# Relic neutrinos and cosmic background radiation: a new way of comparison

P. R. Silva – Retired Associated Professor – Departamento de Física - ICEx – Universidade Federal de Minas Gerais

ABSTRACT – A new equation of state is proposed in order to describe the thermal behavior of relic neutrinos. It is based on extensions of the MIT bag model to deal with the gravitational interaction and takes in account the fermionic character of neutrinos. The results for the temperature and entropy of relic neutrinos are compared with those of the cosmic background radiation, treated as a gas of photons at the temperature of 2.726 K. In particular, it is found that the temperature of the relic neutrinos is ¾ of that of the photon gas. The ratio between the two entropies is also estimate.

1 – INTRODUCTION

   Besides relic photons, neutrinos are one of the most abundant particles in the universe [1]. The number densities of these relic particles (photons and neutrinos) are comparable in magnitude [1]. As was pointed out by Ejnisman and Bigelow [2], an interesting and challenging achievement would be the detection of the relic cosmic neutrino background (CNB). On the other hand, relic photons were detected by Penzias and Wilson [3], through the discovery of the cosmic microwave background (CMB). The temperature of the photon gas of the CMB is known to be today $T_\gamma = 2.726$ K [4]. An experiment of the type-Cosmic Background Explorer (COBE) able do detect the cosmic neutrinos (CNB) should reveal a Fermi-Dirac distribution with a temperature of about 2 K [5]. However until now [6], the temperature of the CNB has not been measured. As was pointed out by Pastor [1], as the universe cools (at energies of about 2 MeV), the weak interaction rate $\Gamma_\nu$ falls below the expansion rate H and neutrinos decouple from rest of the primordial plasma. After decoupling, the neutrinos expand freely exhibiting a Fermi- Dirac distribution. Yet according to Pastor [1], at temperatures of the order of the electron mass, the electron-positron pairs annihilate and transfer their entropy into photons but not in decoupled neutrinos, causing the well-known difference between the temperature of relic photons and relic neutrinos, $T_\gamma / T_\nu = (11/4)^{1/3}$ (please see [7]).
   In this note we are going to compare relic photons and relic neutrinos through different point of view that which is usually considered in the literature. We will propose a new equation of state for the gas of relic neutrinos and then we compare entropy and pressure evaluated from this equation with the those quantities obtained from the equation of state of the relic photon gas. Working with these two equations of state it is also possible to obtain a number for the ratio, $T_\gamma / T_\nu$.

## 2 – A NEW EQUATION OF STATE FOR THE NEUTRINO GAS

Being neutrinos particles having a fermionic character would be interesting to search for a equation of state distinct of that which represents a photon gas (which shows a bosonic signature). An extension of the MIT bag model [8], was proposed by the present author [9] as a means to describe the macrocosmos. The combination of the MIT bag model with another potential which took in account highly relativistic fermionic particles, was used after [10] to evaluate the number and the mass energy of the relic neutrinos.

Therefore by considering references [8,9,10] let us write the equation:

$$P V = \tfrac{1}{4} M c^2. \tag{1}$$

In (1), V is the volume of the observable universe, P is the pressure at its boundary equilibrated by the pressure of the neutrino gas and $M c^2$ is the total mass energy of the matter enclosed by this volume. By identifying this mass with the total energy U of the relic gas of neutrinos, we have

$$P = \tfrac{1}{4} (U/V) = \tfrac{1}{4} u(T), \tag{2}$$

where u(T) is energy density of the relic neutrinos, which is supposed to be a function only of the absolute temperature T. Therefore (2) accounts for the equation of state of the gas of relic neutrinos.

Now, the second law of thermodynamics implies the following relation which holds for all systems

$$(\partial U/\partial V)_T = T (\partial P/\partial T)_V - P. \tag{3}$$

From $PV = \tfrac{1}{4} U$ and the fact that P depends on temperature alone we have

$$(\partial U/\partial V)_T = 4 P = U/V \equiv u(T). \tag{4}$$

Using (3) we have

$$du/u = 5 \, dT/T, \tag{5}$$

and upon integrating we finally obtain

$$u_\upsilon = b_\upsilon T^5. \tag{6}$$

and

$$P_\upsilon = \tfrac{1}{4} b_\upsilon T^5, \tag{7}$$

where $b_\upsilon$ is a constant.

3 – COMPARISON OF RELIC NEUTRINOS AND RELIC PHOTONS

We can suppose that different kinds of particles as neutrinos and photons experiment different pressures of the vacuum at the boundary of the universe. For the relic photon gas we can write

$$P_\gamma = \tfrac{1}{3} b_\gamma T^4. \tag{8}$$

The fact that the two gases are submitted to different vacuum pressures and that each one is separately in thermal equilibrium implies in the attribute of different temperatures for them, namely $T_\upsilon$ and $T_\gamma$.

We may think in a modification of the MIT bag model which leads to the relation

$$P^\# V = \tfrac{1}{3} M c^2. \tag{9}$$

Equation (9) is the same equation of state as the one obeyed by the photon gas. Upon to identify $P^\#$ with $P_\gamma$ and P (of eq. (1)) with $P_\upsilon$, we obtain

$$P_\upsilon / P_\gamma = \tfrac{3}{4}. \tag{10}$$

On the other hand let us think a classical ideal gas with a fixed number of particles and occupying the volume V. Putting this ideal gas (a thermometer) separately in thermal equilibrium with both gases we can write

$$P_\upsilon / P_\gamma = T_\upsilon / T_\gamma = \tfrac{3}{4}. \tag{11}$$

Equation (11) gives

$$T_{\upsilon,now} = \tfrac{3}{4}\, T_{\gamma,now} = 2.045 \text{ K}. \tag{12}$$

The above value for the temperature of relic neutrinos obtained in this work must be compared with $T_{\upsilon,\text{literature}} = 1.946$ K [7].

Relation (11) also implies in

$$b_\upsilon = (T_\gamma / T_\upsilon)^4 \, b_\gamma / T_\upsilon = 1.548 \, b_\gamma / K. \tag{13}$$

Equation (13) can also be obtained by equaling the energy densities of the two gases at the current time (now).

The energy density of relic photons is given by

$$u_\gamma = b_\gamma \, T^4. \tag{14}$$

Comparison of (6) and (14) conceive us to imagine that, due to the different cooling ratios of these two gases in expansion, in the future the two gases will reach the thermal (but not thermodynamical) equilibrium. This will occur at the temperature $T_*$, satisfying the equation

$$b_\upsilon T_*^5 = b_\gamma T_*^4. \tag{15}$$

Solving (15) for $T_*$, taking in account (13), we get

$$T_{\upsilon,\text{future}} = T_{\gamma,\text{future}} = T_* = 0.646 \text{ K}. \tag{16}$$

Now let us estimate the entropy of the gas of relic neutrinos. To do this, we proceed in a way analogous to that worked out in reference [12]. We write

$$dS = (dU + P\, dV)/T = (1/T)[d(uV) + \tfrac{1}{4} u\, dV]$$

or

$$dS = (5/4)\, b_\upsilon \, T^4 \, dV + 5\, V\, b_\upsilon\, T^3\, dT.$$

Integrating the total differential, we find

$$S_\upsilon = (5/4)\, b_\upsilon \, T^4 \, V. \tag{17}$$

In (17), we took the integration constant equal to zero.

On the other hand the entropy of the relic photon gas is given by [12]

$$S_\gamma = (4/3) b_\gamma T^3 V. \tag{18}$$

The ratio of these entropies is then given by

$$S_\upsilon(T_\upsilon)/S_\gamma(T_\gamma) = 5/4 = 1.25. \tag{19}$$

Meanwhile Egan and Lineweaver (EL) [6] have estimate the entropies of these two gases and found

$$S_\gamma|_{EL} = 2.03 \pm 0.15 \times 10^{89} k_B, \tag{20}$$

$$S_\upsilon|_{EL} = 5.16 \pm 0.14 \times 10^{89} k_B. \tag{21}$$

Therefore the ratio between these two entropies estimates by Egan and Lineweaver is greater than that evaluated in the present work by a factor of approximately two.

In the case of the relic neutrino gas, the equation of the adiabat representing the equilibrium expansion is given by

$$T^4 V = \text{constant}. \tag{22}$$

Making use of (17), we can find the heat capacity of relic neutrinos. We have

$$C_V = T (\partial S_\upsilon / \partial T)_V = 5 b_\upsilon T^4 V. \tag{23}$$

Concluding, in this note we propose a new equation of state for relic neutrinos. This equation permit us to compare in thermodynamical grounds some properties of the CMB with the CNB, We remember that relic neutrinos, until now, remains lacking of experimental detection.